\documentclass[aps,prl,twocolumn,groupedaddress]{revtex4-1}
\usepackage{amssymb,amsmath}
\usepackage{color,soul}
\usepackage{graphicx}
\usepackage{hyperref}

\begin{document}

\title{Quasi Long Range Order and Vortex Lattice in the Three State Potts Model}

\author{Soumyadeep Bhattacharya}
\email{sbhtta@imsc.res.in}
\author{Purusattam Ray}
\email{ray@imsc.res.in}
\affiliation{The Institute of Mathematical Sciences,
CIT Campus, Taramani, Chennai 600113, India}

\date{\today}

\begin{abstract}
We show that the order-disorder phase transition in the three
state Potts ferromagnet on a square lattice is driven by a coupled
proliferation of domain walls and vortices.
Raising the vortex core energy above a threshold value decouples
the proliferation and splits the transition into two. 
The phase between the two transitions exhibits an emergent 
U(1) symmetry and quasi long range order.
Lowering the core energy below a threshold value also splits the 
order-disorder transition but the system forms a vortex lattice in the 
intermediate phase.
\end{abstract}


\maketitle

Phase transitions in a variety of systems are driven by the
proliferation of topological defects~\cite{mermin1979topological,
chaikin2000principles, nelson2002defects, vilenkin2000cosmic}.
Manipulation of defects, therefore, provides a natural route
towards altering the nature and location of phase transitions,
which in turn can significantly alter the phase diagram itself.
The role played by a single type of defect, and the effect of
manipulating it, has been studied in models of
superfluids~\cite{kohring1986role,shenoy1990enhancement,zhang1993vortex},
liquid crystals~\cite{lammert1993topology,dutta2004phase}
and Heisenberg ferromagnets~\cite{lau1988role,kamal1993new,
motrunich2004emergent}.
In a large class of systems, however, the phase diagram is
determined by the proliferation of not one but multiple types of
defects~\cite{jose1977renormalization,einhorn1980physical,
strandburg1988two,sadr1999melting,bernard2011two,
chae2012direct,ortiz2012dualities,sato2013quasi}.
We would like to identify a minimal model in which the interplay
between two types of defects, and the effect of manipulating them,
can be studied clearly.

The two state (Ising) ferromagnet on a square lattice is one of
the simplest spin models which exhibit a defect driven phase
transition.
Domain wall defects appear as small loops in the ordered phase
of the model and drive a transition to the disordered phase
upon proliferation~\cite{fradkin1978order, aoki2009domain}.
The next simplest model, the three state Potts ferromagnet, also
exhibits an order-disorder transition~\cite{potts1952some,wu1982potts}.
This model, however, supports the formation of
domain wall as well as $\mathbb{Z}_3$ vortex defects.
An approximate energy versus entropy balance calculation suggests
that the system disorders because vortex-antivortex pairs unbind
as soon as the domain walls proliferate~\cite{einhorn1980physical}.
Apart from this calculation, the role played by the two types of defects
in the model has remained largely unexplored.

\begin{figure}
\includegraphics[width=\linewidth]{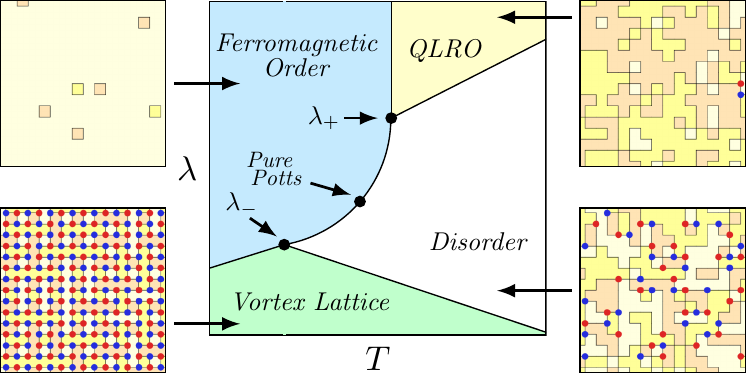}
\caption{\label{fig_phase_diagram}
(Color online) Schematic phase diagram in the parameter
space of temperature $T$ and vortex suppression $\lambda$.
The pure Potts model corresponds to $\lambda=0$. Estimates for
$\lambda_+$ and $\lambda_-$ are given in the text.
Domain walls (black lines), vortices (blue dots) and
antivortices (red dots) are overlaid on typical spin configurations
obtained in each phase.}
\end{figure}

In this Letter, we use Monte Carlo simulations to demonstrate that the
interplay between domain walls
and vortices in the three state Potts model generates a rich phase
diagram [Fig.~\ref{fig_phase_diagram}].
We show that the order-disorder transition in the model is driven by a
coupled proliferation of the two types of defects.
When we raise the core energy of the vortices by an amount $\lambda$,
the model continues to exhibit the order-disorder transition upto a
certain threshold $\lambda = \lambda_+$.
Above $\lambda_+$, the vortices proliferate after the domain walls and
the transition splits into two.
The phase, which appears intermediate between the two transitions, exhibits
an emergent $U(1)$ symmetry and quasi long range order (QLRO).
When we lower the core energy using negative values of $\lambda$,
the order-disorder transition of the pure Potts case
becomes sharper.
Below a threshold $\lambda_-$, the transition again splits into two.
In this case, however, the intermediate phase is a vortex lattice in which
the vortices and antivortices display sublattice ordering.

Before discussing the phase diagram in detail, we describe how the defects
are identified for a given configuration of spins.
Each spin $\sigma_i$, at vertex $i$ on a square lattice
$\Lambda$, can be in one of three states: $\sigma_i \in \{0,1,2\}$.
Domain walls and vortices reside on the edges and vertices, respectively, of
the dual lattice $\Lambda'$ which is a square lattice displaced
from $\Lambda$ by half a lattice spacing along each axis.
If two spins across an edge $\langle i,j \rangle \in \Lambda$ are 
in dissimilar states,
then a domain wall is placed on the dual edge in $\Lambda'$.
The vorticity at each dual vertex $i' \in \Lambda'$ is determined by 
calculating a discrete winding number
$\omega_{i'}$~\cite{ortiz2012dualities}.
For $\mathbb{Z}_3$ vortices,
$\omega_{i'} = (\Delta_{ba} + \Delta_{cb} + \Delta_{dc} + \Delta_{ad})/ 3$
where $\Delta_{ba}$ represents $(\sigma_b - \sigma_a)$ wrapped to lie 
in $[-1,+1]$ 
and $\sigma_a$, $\sigma_b$, $\sigma_c$, $\sigma_d$ are the four
spins on the square plaquette in $\Lambda$ surrounding $i'$ in an 
anticlockwise sense.
A vortex (antivortex) is present at $i'$ if $\omega_{i'}$ 
is $+1$ ($-1$).

\begin{figure*}
\includegraphics[width=\linewidth]{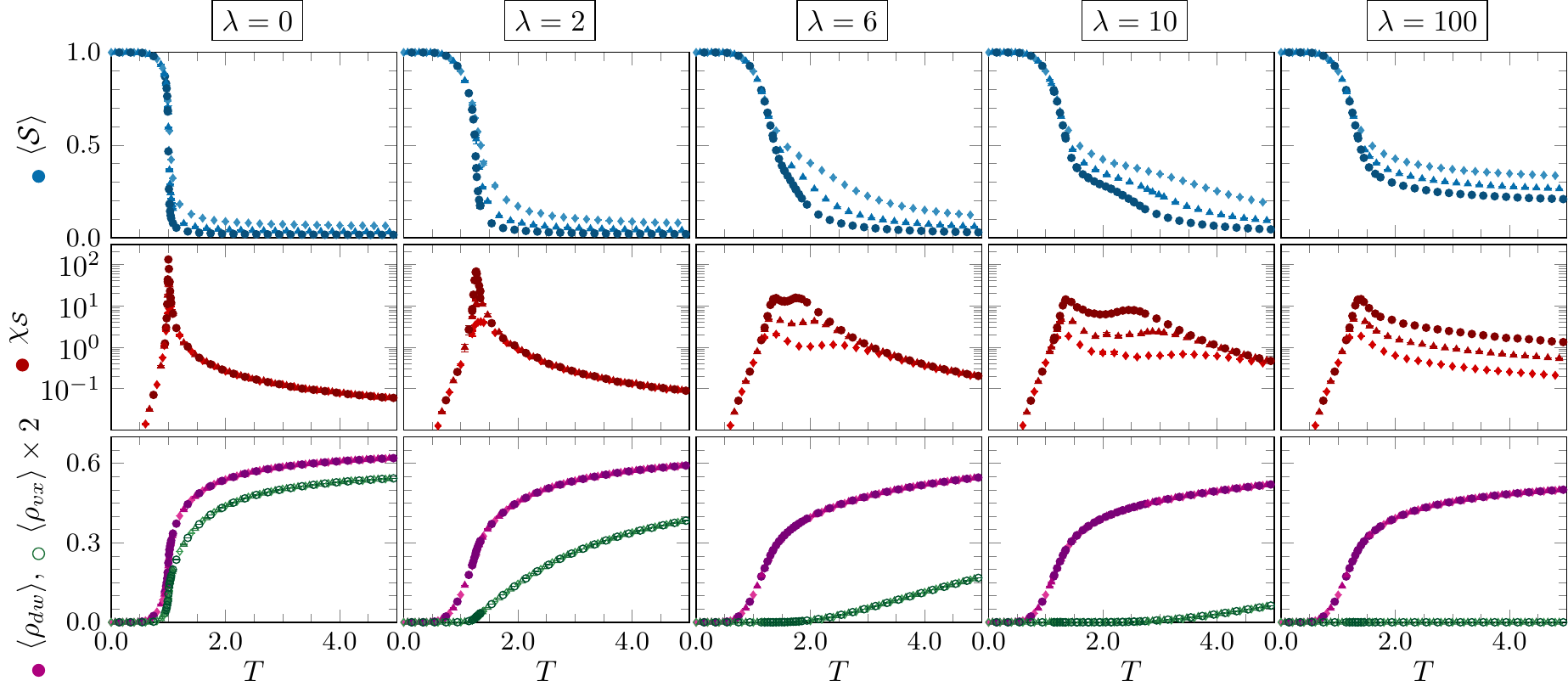}
\caption{\label{fig_lm_plus}
(Color online) The order-disorder transition of the pure Potts model 
($\lambda = 0$), marked by the decay of $\mathcal{S}$ and divergence
of $\chi_{\mathcal{S}}$, is accompanied by a simultaneous increase of
$\rho_{dw}$ and $\rho_{vx}$.
Weak suppression of vortices ($\lambda = 2$) shifts the transition to
a higher temperature. Stronger suppression decouples the simultaneous
proliferation of the defects and splits the transition into two.
The data has been obtained for $L=16$ (diamonds),
$L=32$ (triangles) and $L=64$ (circles).}
\end{figure*}

A standard method for suppressing the formation of vortices in models of
superfluids involves raising the vortex core 
energy by an amount $\lambda$~\cite{kohring1986role,shenoy1990enhancement,
bittner2005vortex,sinha2010role}.
Upon inclusion of such a term for $\mathbb{Z}_3$ vortices, the
three state Potts Hamiltonian, with nearest neighbor ferromagnetic
interaction $J > 0$, becomes
\begin{eqnarray}
\mathcal{H} = J\sum_{\langle i,j \rangle \in \Lambda} 
              (1 - \delta_{\sigma_i,\sigma_j})
              + \lambda \sum_{i' \in \Lambda'} |\omega_{i'}|\,.
\label{eqn_hamiltonian}
\end{eqnarray}
If the number of domain walls and vortex defects corresponding to
a given spin configuration is denoted by $N_{dw}$ and $N_{vx}$, respectively,
then (\ref{eqn_hamiltonian}) can be rewritten as
$\mathcal{H} = J N_{dw} + \lambda N_{vx}$, clearly indicating that the
statistical behavior of the model depends solely on the number of
defects.
In particular, the behavior depends on the number of domain walls and their
three branch intersections.
Each $i' \in \Lambda'$ is visited by zero, two, three or four domain walls.
Out of these four scenarios, $\omega_{i'} \neq 0$ 
only when three domain walls intersect.
The present model can, therefore, be equivalently expressed as a 
domain wall loop model~\cite{dubail2010critical} with a fugacity parameter
$\lambda$ controlling the density of three branch intersections.

We have determined the phase diagram of the model in the two-dimensional
parameter space of $\lambda$ and temperature $T$ by simulating 
(\ref{eqn_hamiltonian}) on a
$L \times L$ square lattice. 
As the plaquette based $\lambda$ term cannot be incorporated into currently
known cluster algorithms, the spins were updated using a single spin-flip
algorithm~\cite{landau2014guide}.
We measured the density of domain walls $\rho_{dw} = N_{dw}/2L^{2}$, 
the density of vortices $\rho_{vx} = N_{vx}/L^{2}$
and the Potts order parameter 
$\mathcal{S} = 3(\max\{n_0,n_1,n_2\} - 1/3)/2$,
where $n_\sigma$ represents the fraction of spins in state $\sigma$.
Large autocorrelation times, arising from the use of the single spin-flip
algorithm, were estimated for these observables and
measurements were made over $10^5 - 10^6$ uncorrelated configurations.

In the pure Potts case ($\lambda = 0$), the order 
parameter decays and the susceptibility
$\chi_{\mathcal{S}} = L^2(\langle \mathcal{S}^2 \rangle - \langle \mathcal{S} \rangle^2)/T$
diverges [Fig.~\ref{fig_lm_plus}] close to the transition
temperature $T_c = 1/\log(1+\sqrt{3}) = 0.9949$~\cite{wu1982potts}.
The transition is accompanied by a simultaneous proliferation of both
types of defects, as indicated by an increase in their densities.
The transition and defect proliferation shift to a higher temperature
$T \approx 1.26$ when vortices are weakly suppressed using $\lambda = 2$
[Fig.~\ref{fig_lm_plus}].
The location of the transition continues to shift in this manner, with
increasing $\lambda$, upto a threshold value
$\lambda = \lambda_+$ which we estimate to lie around $\lambda_+ \approx 8$.

Above $\lambda_+$, the order parameter clearly exhibits a two step 
decay and the susceptibility shows two distinct peaks [Fig.~\ref{fig_lm_plus}].
For $\lambda = 10$, the first decay from the ordered phase to the intermediate
phase occurs around $T \approx 1.4$ and is accompanied by the proliferation
of domain walls.
The vortices proliferate near the second decay, which marks the transition 
from the intermediate phase to the disordered phase at $T \approx 2.7$.
Extremely strong suppression of vortices keeps the first decay unchanged
but shifts the second decay to $T \rightarrow \infty$, thus establishing the
role of vortices in driving the disordering transition.

Two step decay of magnetization, driven by successive proliferation
of domain walls and vortices, has been discussed in the context of 
$\mathbb{Z}_n$ vector Potts (clock) spin models with 
$n \geq 5$~\cite{jose1977renormalization,elitzur1979phase,einhorn1980physical,
frohlich1982massless,lapilli2006universality,baek2009true,van2011discrete,
ortiz2012dualities,borisenko2012phase}.
These models exhibit a phase, intermediate between order and disorder, where
domain walls proliferate but vortices do not.
In the intermediate phase, the system fragments into numerous domains in a
manner such that the spins fluctuate by arbitrary amounts over large 
distances and exhibit a $U(1)$ symmetry upon 
coarse-graining~\cite{einhorn1980physical}.
The emergent continuous symmetry destroys long range order and gives rise
to a quasi long range order which is characterized by a power law decay of 
two-point correlation
$\mathcal{C}(r) = \langle \cos(2 \pi (\sigma_0 - \sigma_r)/3) \rangle \propto r^{-\eta}$,
where $\sigma_0$ and $\sigma_r$ are spins located at a Eucledian distance $r$ apart on
the lattice.
The exponent $\eta$ changes continuously with temperature 
throughout the quasi long range ordered phase until vortex proliferation 
disorders the system~\cite{ortiz2012dualities}.

\begin{figure}
\includegraphics[width=\linewidth]{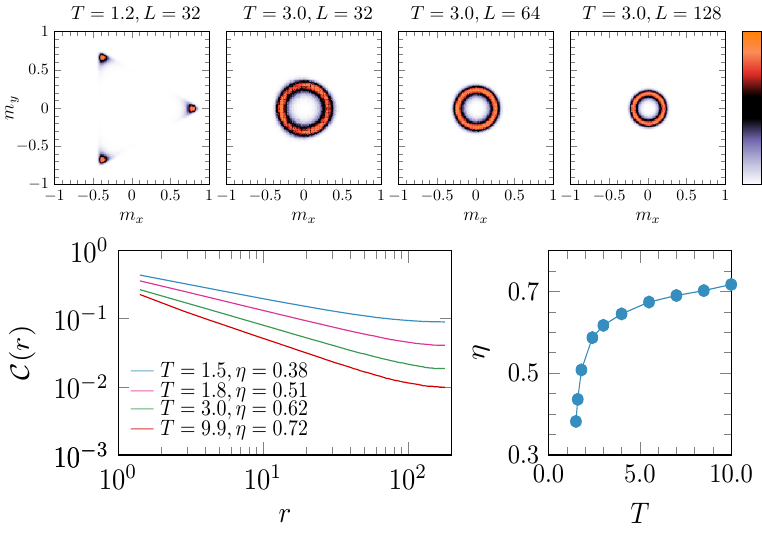}
\caption{\label{fig_l100_props}
(Color online) Top panel shows $P(m_x,m_y)$ in the ordered ($T = 1.2$) and quasi long
range ordered ($T=3.0$) phases for $\lambda = 100$.
Bottom panel shows power-law decay of $\mathcal{C}(r)$ at different temperatures
in the latter phase for $L=256$. The saturation at large $r$ is due
to the finite size of the system.
$\eta$ increases with temperature as shown on the right.
}
\end{figure}

We have measured the distribution $P(m_x,m_y)$ of the $\mathbb{Z}_3$ order
parameter $m = m_x + i m_y$, where
$m_x = \sum_\sigma n_\sigma \cos(2 \pi \sigma/3)$ and 
$m_y = \sum_\sigma n_\sigma \sin(2 \pi \sigma/3)$~\cite{
baek2009true,borisenko2012phase}.
The distribution [Fig.~\ref{fig_l100_props}] clearly indicates a breaking
of the three-fold symmetry in the ordered phase and an enhancement to 
$U(1)$ symmetry in the intermediate phase.
The $U(1)$ symmetry survives an increase in system size while the magnetization
$(m_x^2+m_y^2)^{1/2}$ tends to zero in accordance with the Mermin-Wagner
theorem~\cite{mermin1966absence,archambault1997magnetic}.
$\mathcal{C}(r)$ exhibits a power-law decay throughout the intermediate phase.
$\eta$ increases with temperature from $\eta \approx 0.35$ and appears to
saturate around $\eta \approx 0.75$ at high temperatures
[Fig.~\ref{fig_l100_props}].
This confirms that the intermediate phase is indeed a quasi long range
ordered phase.

\begin{figure*}
\includegraphics[width=\linewidth]{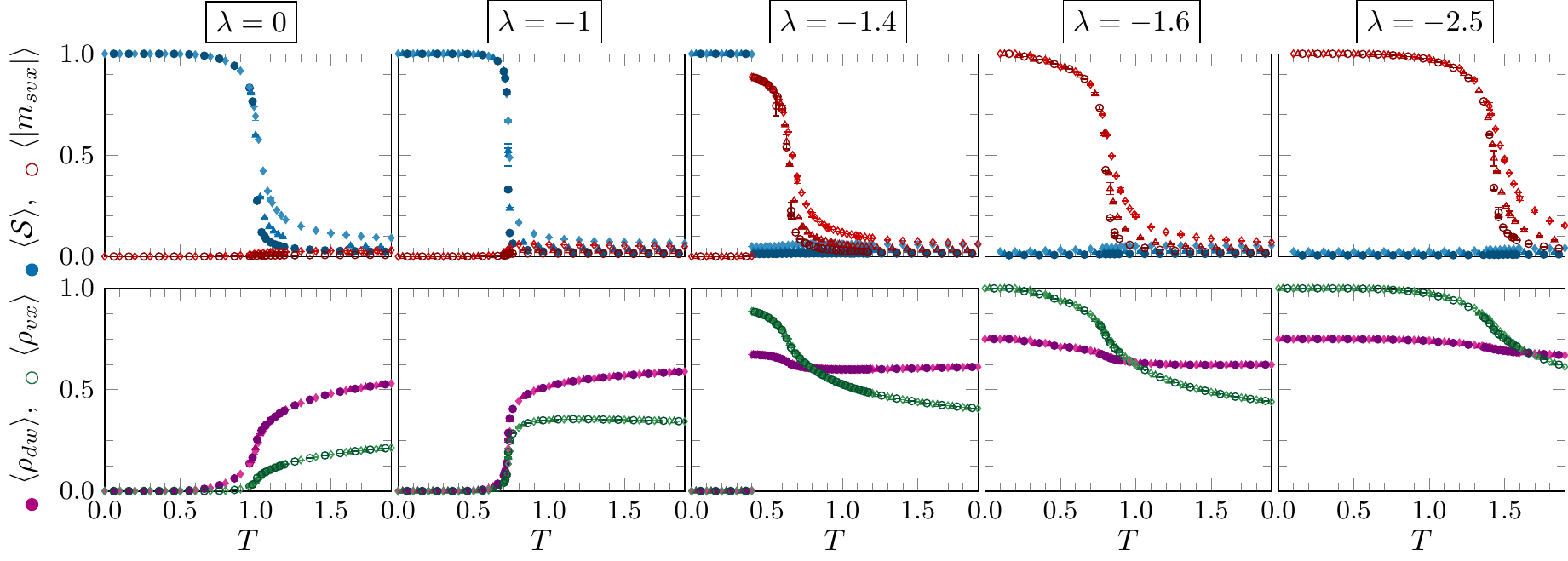}
\caption{\label{fig_lm_minus}
(Color online) Weak enhancement of vortices ($\lambda = -1$) 
results in a sharp decay of $\mathcal{S}$ accompanied by simultaneous 
increase of $\rho_{dw}$ and $\rho_{vx}$. Stronger enhancement ($\lambda = -1.4$)
opens up an intermediate vortex lattice phase characterized by a non-zero value
of $\langle |m_{svx}| \rangle$. With further enhancement, the ordered phase
vanishes while the vortex lattice melts at a higher temperature. Data corresponds
to system sizes mentioned in Fig.~\ref{fig_lm_plus}.}
\end{figure*}

We now turn to the regime $\lambda < 0$. The formation of vortices is
enhanced in this regime and the order-disorder transition of the pure Potts
case shifts to lower temperatures when $\lambda$ is made more 
negative [Fig.~\ref{fig_lm_minus}].
Additionally, the decay of the order parameter grows sharper and the
simultaneous rise in the densities of domain walls and
vortices across the transition becomes more abrupt compared to the pure Potts
case.
When $\lambda$ is decreased below a threshold value $\lambda_- \approx -1.3$,
the model exhibits three distinct regimes [Fig.~\ref{fig_lm_minus}].
The densities of domain walls and vortices are nearly zero in the
ordered phase, show a sharp jump to a large value at intermediate temperatures
and decrease gradually in the disordered phase.
The order parameter, on the other hand, shows a sharp decay from the ordered
phase to the intermediate regime, and remains zero thereafter.
This might suggest disordered behavior in the intermediate regime but an
inspection of spin configurations in that 
regime [Fig.~\ref{fig_phase_diagram}] reveals that the spins are not 
disordered.
Instead, they exhibit a weave pattern which corresponds to an ordering of
the vortex defects: vortices reside on one sublattice and antivortices 
reside on the other.

In order to characterize the sublattice ordering of the vortices,
we define a variable $\epsilon_{i'}$ which is either +1 or -1 depending on the
sublattice of the dual vertex $i'$.
Since some of the dual vertices are vacanct ($\omega_{i'} = 0$),
the sublattice vortex order parameter can be chosen to be of the same form 
as that for a site-diluted Ising antiferromagnet~\cite{kim1985hysteretic,lidmar1997monte}:
$m_{svx} = L^{-2}\sum_{i' \in \Lambda'} \epsilon_{i'} \omega_{i'}$.
The magnitude of this order parameter becomes non-zero in the vortex
lattice and clearly demarcates the intermediate phase from the 
ordered and disordered phases [Fig.~\ref{fig_lm_minus}].
Since this order parameter, based on topological defects, is able to 
distinguish the vortex lattice phase from the disordered phase,
while the symmetry based order parameter $\mathcal{S}$ is unable to do so,
the vortex lattice phase provides a simple example of classical 
topological order.
Formation of vortex lattices in superfluids and superconductors has been extensively studied
over the past few decades~\cite{zhang1993vortex,lidmar1997monte,hu1993two,
gabay1993vortex,blatter1994vortices}
In superconducting thin films, proliferation of dislocations and disclinations
drive a two step structural melting of the vortex lattice~\cite{strandburg1988two,
guillamon2009direct,zehetmayer2015vortex,ganguli2015disordering}.
In the present model, the melting of the vortex lattice to the
disordered phase occurs via a single step process as indicated
by the decay of $\langle |m_{svx}| \rangle$ [Fig.~\ref{fig_lm_minus}].
The sublimation of the vortex lattice to the ordered phase,
which occurs at a lower temperature, is marked by a sharp decay of 
$\langle |m_{svx}| \rangle$.
With a further decrease of $\lambda$, the melting shifts
to higher temperatures [Fig.~\ref{fig_lm_minus}].
The sublimation, on the other hand, shifts to lower temperatures and
hits the zero temperature limit at $\lambda \approx -1.5$.
Below this $\lambda$, the ordered phase is absent and the model exhibits 
a single melting transition from the vortex lattice to the disordered phase.

The nature of the new phases uncovered above is quite clear from the data
obtained for small systems.
A precise estimate of the temperature and nature of the transitions, on the
other hand, requires detailed analysis of data from large systems and will 
be presented separately.
Here, we make a few comments regarding the possible nature of the transitions.
The order-disorder transition for $\lambda = 0$ is of the second order 
type~\cite{wu1982potts,baxter2007exactly}.
This order-disorder transition continues to occur throughout the range 
$\lambda_- < \lambda < \lambda_+$ and is accompanied by a coupled proliferation
of domain walls and vortices.
However, the decay of the order parameter appears to grow weaker with
increasing $\lambda$ [Fig.~\ref{fig_lm_plus}] and sharper with decreasing
$\lambda$ [Fig.~\ref{fig_lm_minus}].
This observation leads us to conjecture that the critical exponents of the transition 
might vary with $\lambda$.
The four state (Ashkin-Teller) ferromagnet is known to exhibit continuously
varying criticality along a line of order-disorder transitions due to
an interplay between vortices and domain walls~\cite{kadanoff1979multicritical}.
The effect of the coupled proliferation on the nature of the transition
in the present model, therefore, seems to be an interesting problem.
In this context, we note that claims of continuously varying criticality 
in the closely related three state chiral Potts model continues to be a 
controversial issue~\cite{huse1982domain,centen1982non,sato2000numerical}.

For $\lambda > \lambda_+$, the order-disorder transition
splits into two and the intermediate phase exhibits quasi long range order.
In $\mathbb{Z}_n$ ferromagnets with $n \geq 5$, the two transitions
bordering the quasi long range ordered phase are known to be of the
Berezinskii-Kosterlitz-Thouless type~\cite{ortiz2012dualities,
borisenko2012phase}.
We can expect the two transitions in the present model to 
be of that type as well.
In $\mathbb{Z}_n$ models, the decay exponent $\eta$ changes from
$\eta = 4/n^2$ at the low temperature transition to $\eta=1/4$ at the
high temperature transition~\cite{borisenko2012phase}.
The values of $\eta \in (0.35,0.75)$ obtained in the quasi long range
ordered phase of the present model [Fig.~\ref{fig_l100_props}] fall beyond
that range.
This deviation from the standard $\mathbb{Z}_n$ model scenario indicates that
the bounds for $\eta$ can change with vortex suppression.

Our estimate of $\lambda_+ \approx 8$ is an approximate one.
For $\lambda = 6$, the intermediate phase is narrow in small 
systems and appears to shrink with increasing $L$ [Fig.~\ref{fig_lm_plus}].
On the other hand, for fixed $L$, the extent of the phase increases with
with increasing $\lambda$.
This competing effect of $L$ and $\lambda$ offers two
possibilites: (a) there exists a threshold $\lambda_+$, above which the 
intermediate region has a non-zero extent in the thermodynamic
limit, or (b) the intermediate region shrinks to a point in the thermodynamic
limit for all finite $\lambda$ and the model exhibits an extended quasi long 
range ordered phase only in the $\lambda \rightarrow \infty$ limit.
A precise estimate of $\lambda_+$, therefore, remains an open problem and
is reminiscent of the problem regarding the location of the Lifshitz point
in the three state chiral Potts 
model~\cite{howes1983quantum,haldane1983phase,schulz1983phase,von1984finite,duxbury1984wavevector}.

In the range $\lambda_- < \lambda < 0$, the order-disorder transition
grows sharper with decreasing $\lambda$, hinting at the possibility that
the transition will become discontinuous before $\lambda$ goes below 
$\lambda_-$.
Such a scenario is quite plausible because an abrupt proliferation of
vortices, similar to the behavior shown in Fig.~\ref{fig_lm_minus}, 
is known to induce discontinuous 
behavior~\cite{van1984continuous,sinha2010role}.
For $\lambda < \lambda_-$, the transition from the ordered phase to the
vortex lattice phase [Fig.~\ref{fig_lm_minus}] is clearly discontinuous.
The gradual decay of $\langle |m_{svx}| \rangle$ between the vortex lattice
phase and the disordered phase, on the other hand, suggests that the melting
transition might be second order in nature. 

The universality class of the three state Potts transition has a ubiquitous
presence in the physics of statistical~\cite{wu1982potts,binder1982theoretical,
freimuth1985phases,hwang1988thermal,baxter1980hard,van2012phase,
szabo2001phase,szolnoki2005three,takaishi2005simulations,zhao2013kosterlitz}, 
quantum~\cite{boninsegni2005supersolid,lecheminant2012exotic,zhuang2015phase,
suzuki2015thermal,rapp2006dynamical} and 
gauge systems~\cite{savit1980duality,
yaffe1982,svetitsky1982,lepori2009particle,borisenko2012phase}.
We have shown that the transition is driven by a coupled
proliferation of domain walls and vortices.
By manipulating the formation of the defects, we have uncovered two new phases 
in the model.
Apart from the exciting possibility that these phases might be
realizable in some of the systems, our work provides a step towards
better understanding the role of topological defects and the presence
of topolgical order in classical spin models.

The authors thank Rajesh Ravindran, Ronojoy Adhikari, Deepak Dhar, Kedar Damle
and Gautam Menon for helpful discussions.

\end{document}